\documentclass{IEEEtran4PSCC}
\usepackage[utf8]{inputenc}
\usepackage[T1]{fontenc}

\usepackage[cmex10]{amsmath}
\usepackage{graphicx}
\usepackage{multirow}
\usepackage{xcolor}
\usepackage{eurosym}
 \usepackage{cite}
 
 \usepackage{tikz}
\usepackage{xcolor, colortbl}
\usetikzlibrary{decorations.pathreplacing}
\usetikzlibrary{arrows,shapes,positioning}
\usetikzlibrary{patterns}
\usepackage{pgfplots}
\pgfplotsset{compat=newest}
\pgfplotsset{plot coordinates/math parser=false}
\pgfkeys{/pgf/number format/.cd,1000 sep={\,}}
\usetikzlibrary{plotmarks}

\newlength\matlabfigurewidth
\makeatletter
\let\old@ps@headings\ps@headings
\let\old@ps@IEEEtitlepagestyle\ps@IEEEtitlepagestyle
\def\psccfooter#1{%
    \def\ps@headings{%
        \old@ps@headings%
        \def\@oddfoot{\strut\hfill#1\hfill\strut}%
        \def\@evenfoot{\strut\hfill#1\hfill\strut}%
    }%
    \def\ps@IEEEtitlepagestyle{%
        \old@ps@IEEEtitlepagestyle%
        \def\@oddfoot{\strut\hfill#1\hfill\strut}%
        \def\@evenfoot{\strut\hfill#1\hfill\strut}%
    }%
    \ps@headings%
}
\makeatother

\psccfooter{%
        \parbox{\textwidth}{\hrulefill \\ \small{21st Power Systems Computation Conference} \hfill \begin{minipage}{0.2\textwidth}\centering \vspace*{4pt} \includegraphics[scale=0.06]{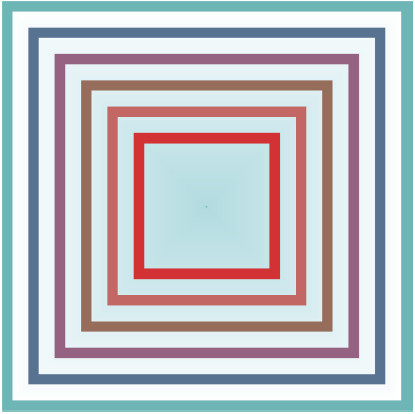}\\\small{PSCC 2020} \end{minipage} \hfill \small{Porto, Portugal --- June 29 -- July 3, 2020}}%
}

\begin{document}

\title{Siting and Sizing of Energy Storage Systems: Towards a Unified Approach for Transmission and Distribution System Operators for Reserve Provision and Grid Support}

% Former title: A Unified Approach for Transmission and Distribution System Operators to Plan Energy Storage for Reserve Provision and Grid Support

%% To specify the authors when (number of affiliations <= 2)
\author{
\IEEEauthorblockN{Stefano Massucco\IEEEauthorrefmark{1}, Paola Pongiglione\IEEEauthorrefmark{1}, Federico Silvestro\IEEEauthorrefmark{1}, Mario Paolone\IEEEauthorrefmark{2}, Fabrizio Sossan\IEEEauthorrefmark{3}}
\IEEEauthorblockA{\IEEEauthorrefmark{1}Universit\'a di Genova - Dipartimento di Ingegneria Navale, Elettrica, Elettronica e delle Telecomunicazioni (DITEN)} \IEEEauthorblockA{\IEEEauthorrefmark{2}Swiss Federal Institute of Technology of Lausanne (EPFL) - Distributed Electrical Systems Laboratory (DESL)}
\IEEEauthorblockA{\IEEEauthorrefmark{3}Mines ParisTech - Centre Procédés, Energies Renouvelables
et Systèmes Energétiques}
%\and
%\IEEEauthorblockN{Author n.1 Name per Affiliation B\\ Author n.2 Name per Affiliation B}
%\IEEEauthorblockA{(Affiliation B) Department Name of Organization \\
%Name of the organization, acronyms acceptable\\
%City, Country\\
%\{email author n.1, email author n.2\}@domain (if desired)}
}

%% To specify the authors when (number of affiliations > 2)
% \author{\IEEEauthorblockN{Author n.1\IEEEauthorrefmark{1},
% Author n.2\IEEEauthorrefmark{2},
% Author n.3\IEEEauthorrefmark{3}, 
% Author n.4\IEEEauthorrefmark{3} and
% Author n.5\IEEEauthorrefmark{4}}
% \IEEEauthorblockA{\IEEEauthorrefmark{1} Department Name of Organization A\\
% Name of the organization A,
% Address A\\ Emails if wanted}
% \IEEEauthorblockA{\IEEEauthorrefmark{2} Department Name of Organization B\\
% Name of the organization B,
% Address B\\ Emails if wanted}
% \IEEEauthorblockA{\IEEEauthorrefmark{3} Department Name of Organization C\\
% Name of the organization C,
% Address C\\ Emails if wanted}
% \IEEEauthorblockA{\IEEEauthorrefmark{4}Department Name of Organization D\\
% Name of the organization D,
% Address D\\ Emails if wanted}
% }

\maketitle

\begin{abstract}
%This paper presents an algorithm for vertical and horizontal planning of storage, that aims at determining the optimal locations, energy capacity and power rating of distributed storage at multiple voltage levels accounting for the needs of both distribution and transmission systems.  The problem is formulated as an optimal power flow that minimizes the total cost of operation (provision of conventional reserve + capital investment for ESSs) while subject to operational constraints of DSOs’and TSO’s grids, grid losses, \textbf{and efficiency of ESSs}. Performance is validated by simulation considering the HV 9-bus IEEE test system, 2 CIGRE’ MV test systems, a fleet of conventional generators and distributed generation, under several scenarios that depict deviations of stochastic renewable generation from day-ahead forecasts.
This paper presents a method to determine the optimal location, energy capacity, and power rating of distributed battery energy storage systems at multiple voltage levels to accomplish grid control and reserve provision. We model operational scenarios at a one-hour resolution, where deviations of stochastic loads and renewable generation (modeled through scenarios) from a day-ahead unit commitment and violations of grid constraints are compensated by either dispatchable power plants (conventional reserves) or injections from battery energy storage systems. By plugging-in costs of conventional reserves and capital costs of converter power ratings and energy storage capacity, the model is able to derive requirements for storage deployment that achieve the technical-economical optimum of the problem. The method leverages an efficient linearized formulation of the grid constraints of both the HV (High Voltage) and MV (Medium Voltage) grids while still retaining fundamental modeling aspects of the power system (such as transmission losses, effect of reactive power, OLTC at the MV/HV interface, unideal efficiency of battery energy storage systems) and models of conventional generator. A proof-of-concept by simulations is provided with the IEEE 9-bus system coupled with the CIGRE’ benchmark system for MV grids, realistic costs of power reserves, active power rating and energy capacity of batteries, and load and renewable generation profile from real measurements.

\end{abstract}


\section{Introduction}
The use of energy storage systems (ESSs) has been advocated to cope with the intermittency of distributed stochastic renewable generation and mitigate its impact on operational practices of transmission system operators (TSOs) and distribution system operators (DSOs). Proposed applications of ESSs range from energy arbitrage, support to primary frequency control and reserve provision for services at the system level, up to grid control, congestion management, dispatch of local systems and self-consumption for services at the level of distribution systems \cite{koller2015review, OULDAMROUCHE201620914, 7542590,7429781,8464265}. It is also well understood that ESSs can be used to provide multiple grid services, leading to increased exploitation of their energy capacity and power rating, and a shorter return on investment, thus making relevant to plan ESSs accounting for multiple services\cite{STAFFELL2016212, 7422904, wu2015energy, 7932132, Namor_tsg_2018}. A fundamental question for grid operators considering ESSs is how to determine their size (i.e., energy capacity and apparent power rating) and location. This problem has been extensively investigated in the existing technical literature (see, e.g., \cite{zidar2016review}), with methods ranging from optimal power flows (OPFs) \cite{Mostafa_ESSplan} to heuristics \cite{Pandzic, CELLI2018154}. More tractable linearized grid models are also widely, as in \cite{fortenbacher2016optimal} for the siting and sizing problem and in \cite{8918388, Powertech} for volt/var control.

Most of the existing literature focuses on a single voltage level (i.e., transmission grid or distribution grid) and a single set of services at a time, which are specific to TSOs or DSOs only. In this paper, we propose a modeling framework to determine the optimal location, energy capacity and power rating of distributed battery energy storage systems accounting for multiple voltage levels simultaneously and modeling the provision of ancillary services to both a DSO and TSO. 
%OLD: Especially, the ancillary services that we consider are grid control (i.e., voltage control and congestions management) and reserve provision..
%MARIO better differentiate between obj fnc and problem's constraints
We refer to this formulation as vertical and horizontal planning of ESSs, as opposed to the works discussed above that consider a single voltage level and a single class of services at a time (horizontal planning). Especially, we consider the case of voltage control and congestion management in distribution grids, and the provision of regulating power to the TSO. The work in \cite{8302948} tackles a problem similar to ours. It considers the joint ESSs siting and sizing problem in distribution and transmission grids (modeled with a SOCP-based OPF and DC load flow, respectively). Its objectives are, for DSOs, maximizing the revenue from energy trading and achieving local grid control and, for the TSO, maximizing the social welfare by minimizing the electricity price in a wholesale market. This paper approaches the same problem with a different perspective: we consider intra-day reserve procurement and perform an economic evaluation between when it is provided by battery-based ESSs versus conventional generation. %This stands as the main contribution of this paper.
The proposed decision problem relies on a linear optimal power flow model, that computes the nodal injections (from ESSs or conventional power plants) to compensate for variations with respect to a day-ahead unit commitment by minimising an economical cost and while subject to grid constraints. The economic cost is the sum of the operational costs of activating reserves from conventional power plants and the capital costs to install ESSs (i.e., apparent power rating and energy storage capacity). As energy storage devices, we consider lithium-ion batteries, that are modeled as lossy bulk energy reservoirs (i.e., charge/discharge efficiency is taken into account), and limits due to energy storage capacity and apparent power rating of the converters. Grid constraints on nodal voltages, cables ampacities, and apparent power flows at the grid connection point are modeled with linearized grid models that, while trading some accuracy, allow for an efficient (convex) formulation of the problem and retain the possibility of modeling important operational aspects, like the effects of reactive power and losses, and on-load tap changer (OLTC) transformers.
A proof-of-concept of the proposed planning algorithm is delivered by simulations considering a case study with both HV and MV grids (modeled according to the specifications of the IEEE 9-bus test system and CIGRE’ benchmark system for MV networks) equipped with conventional power plants, demand, and distributed photovoltaic (PV) generation. These lasts are modeled using measurements from real systems.

The rest of the paper is organized as follows: Section II describes the problem to be solved, Section III introduces the adopted formulation and modeling solutions, Section IV presents the study case and simulation scenarios used to test the proposed algorithm. Section V introduces and discusses the results and, finally, Section VI draws the conclusions and future developments.

\section{Problem Statement} % ok
We refer to the case study shown in Fig.~1, that illustrates a power system with a meshed HV transmission grid and a MV distribution grid.
The HV system interfaces conventional power plants G1-G3 through step-up transformers. At HV bus 9, two OLTC transformers feed two MV networks with demand and distributed renewable generation. HV nodes 5 and 7 feature aggregated demand and renewable generation from downstream networks, modeled in a lumped way as nodal apparent power injections. HV node 1 is the slack bus of the system. The system topology is fixed, so we do not model possible operational topological changes (see, e.g., \cite{Mostafa_ESSplan}). The problem is determining the optimal location in the system (both at the HV and MV levels) and specifications (energy capacity and power rating of the converter) of battery energy storage systems.

\begin{figure}[!ht]
    \centering
    \includegraphics[width=0.99\columnwidth]{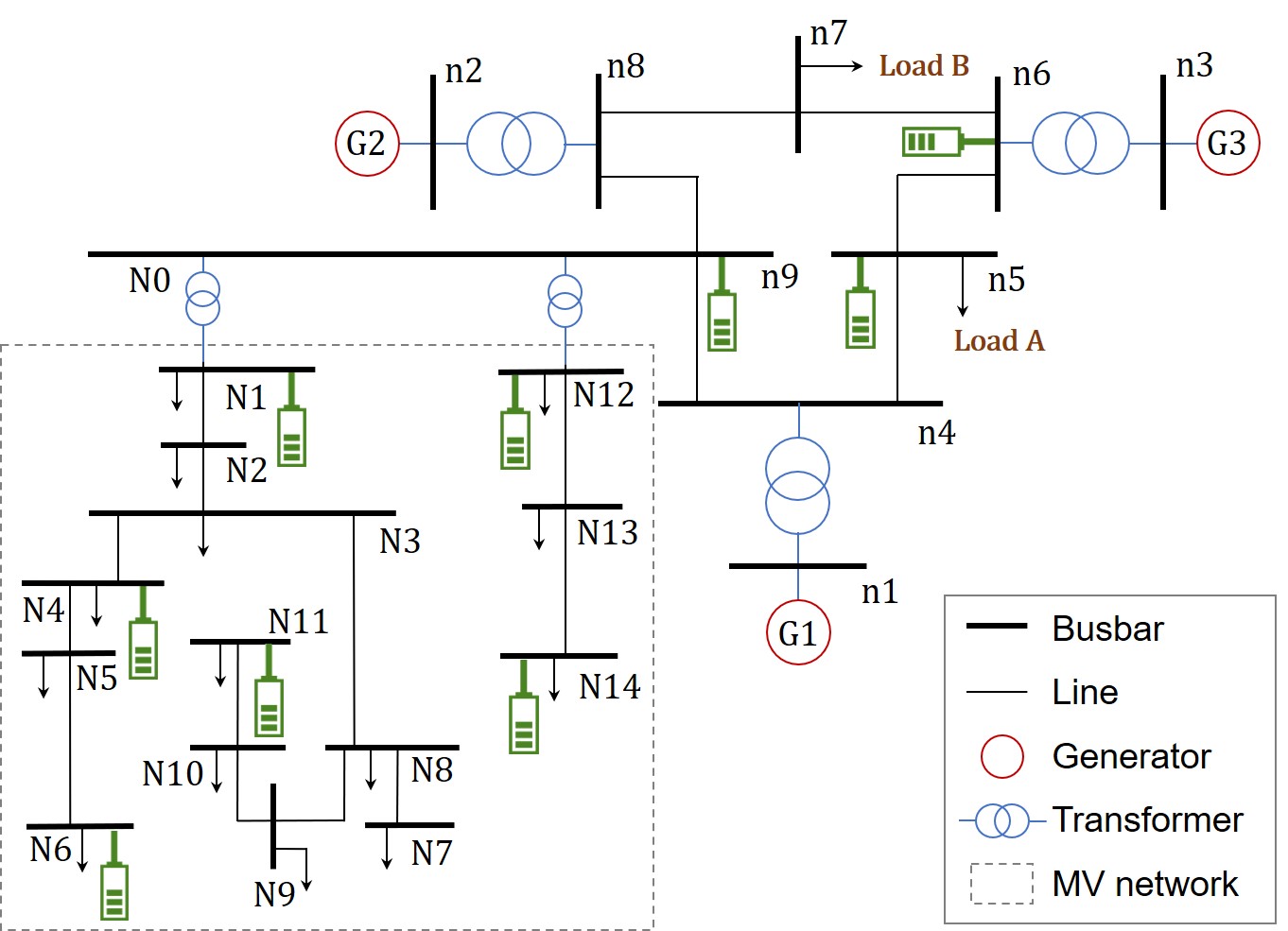}
    \caption{The case study with an HV and MV grids, conventional generators, loads, and distributed PV generation (not shown). The green icon denotes the location of the batteries (step-up transformers are omitted for simplicity), that is an output of the problem together with their power rating and energy capacity requirements. Load A and Load B refer to aggregated injections of other MV networks.}
    \label{fig:Case_Study}
\end{figure}

The operational paradigm that we model and exploit to determine ESSs locations and specifications is the power balance with an one-hour sampling. We first assume that a unit commitment process performed  in the day-ahead stage determines the schedule of the conventional generation units and tap position of OLTC transformers. This achieves the (active) power balance in the system according to day-ahead forecasts of the aggregated nodal injections at the HV level as a function of the generation costs and subject to the constraints of the transmission capacity. Then, in real-time, the realization of the stochastic demand and renewable generation might vary from day-ahead forecasts. This determines a power imbalance in the system and deviations from the day-ahead plan with, possibly, new violations of network constraints. The power balance mismatch and network violations are compensated for and solved by activating power reserves from conventional generation unit or injection from ESSs by solving an optimal inter-temporal power flow problem. This mechanism is the base of the siting and sizing problem proposed in this paper and is thoroughly described in the next section. Thanks to assigning operational costs to the activated reserve from conventional power plants and capital costs for the installation of ESSs (i.e., connection costs, energy capacity costs, and converter power rating costs), the algorithm determines an optimal economic trade-off between the reserve to deploy from conventional units and installed energy storage capacity and power rating. We leverage different scenarios of the stochastic demand and distributed generation to derive siting and sizing guidelines that account for multiple realizations of the uncertain elements.

At this stage, we consider power reserves for secondary regulation, and we do not model the grid frequency.

%Even when the provision of services to the upper-layer grid is considered (e.g., energy arbitrage through time-of-use tariffs or reserve provision through broadcasting market prices), a price-taker approach is typically adopted [---], thus disregarding closed-loop interactions with the system that is essential to asses. %that could lead to non-scalable and erroneous planning guidelines. 
%%% TO DO Paola

\section{Methods}
We now describe the foundational models used in the formulation and, as the last element, the siting and sizing problem.

\subsection{Grid model}
We implement a grid model to determine the nodal injections and position of the OLTCs' tap such that the grid constraints on voltage magnitudes and line ampacities are respected, while accounting for losses in the power lines. Load flows are notoriously non-linear, and their inclusion in optimization problems leads to non-convex formulations. For tractability, we model grid constraints with linear functions using the notion of sensitivity coefficients (SCs). When solving the load flow with the Newton-Raphson method, the SCs of the voltage magnitudes can be derived easily by extracting the proper submatrix from the inverse Jacobian at the last iteration of the algorithm. In this paper, we compute SCs with the method described in \cite{christakou2013efficient} that derives the SCs of the voltage magnitudes not only against the active and reactive power injections but also the slack voltage. As illustrated in \cite{christakou2013efficient}, SCs are determined by solving a system of linear equations as a function of the grid's admittance matrix and voltage and currents phasors at a working point.\footnote{The linear system has a unique solution when the load-flow Jacobian is locally invertible \cite{paolone2015optimal}.} Voltage and current phasors at the working point are determined offline with a load flow as a function of the nominal nodal injections, which correspond to the same demand and distributed generation day-ahead point predictions used to determine the day-ahead schedule of the power plants (similarly to \cite{guptal2019performance}). Let $t$ denote the time index, $\boldsymbol{v_t}$ and $\boldsymbol{i_t}$ be vectors collecting all nodal voltage magnitudes and line currents of a given network at time $t$, and $\boldsymbol{P_t}$ and $\boldsymbol{Q_t}$ nodal active and reactive power injections, and $v_{0t}$ the voltage magnitude of the grid's slack bus. Nodal voltages and line currents are:
\begin{subequations} \label{eq:gridmodel}
\begin{align}
& \boldsymbol{v}_t = K_{vt} \begin{bmatrix} \boldsymbol{P_t} & \boldsymbol{Q_t} \end{bmatrix}^T + v_{0t} \boldsymbol{b_t}   + \boldsymbol{c_t} \label{eq:gridmodel0}\\
& \boldsymbol{i}_t = K_{it} \begin{bmatrix} \boldsymbol{P_t} & \boldsymbol{Q_t} \end{bmatrix}^T + v_{0t} \boldsymbol{d_t} + \boldsymbol{e_t} \label{eq:gridmodel1}
\end{align}

where matrices $K_{vt}, K_{it}$ and vectors $\boldsymbol{b_t}, \boldsymbol{c_t}, \boldsymbol{d_t}, \boldsymbol{e_t}$ are  derived from the so-called sensitivity coefficients \cite{christakou2013efficient}. The active and reactive power flow at the slack bus of the grid $P_{0t}, Q_{0t}$ at time $t$ is modeled as
\begin{align}
& \begin{bmatrix} P_{0t} & Q_{0t} \end{bmatrix}^T = \boldsymbol{K_{St}} \begin{bmatrix} \boldsymbol{P_t} & \boldsymbol{Q_t} \end{bmatrix}^T + v_{0t} f_t + g_t \label{eq:gridmodel2}
\end{align}
\end{subequations}
where $\boldsymbol{K_{St}}$ is a vector, and $f_t$ and $g_t$ are scalar. They are also derived from the sensitivity coefficients.

All the (time varying) parameters of linear models \eqref{eq:gridmodel0}-\eqref{eq:gridmodel2} depend on respective linearization points, that, in this work, correspond to the day-ahead forecasts of the nodal injections. Though depending on the state, $K_{vt}$ and $K_{it}$ are supposed to change smoothly with respect to state variations.

\subsection{Grid continuity constraints}
%For the reason that it eill be evident soon,
The linear grid models (1) of the HV and MV network are derived separately for each grid, i.e., without integrating all the grid information into a single admittance matrix. For a coherent representation of the whole system and obeying to conservation principles, we need continuity constraints on the apparent power flow and voltage at the MV-to-HV interface. Let $b$ the bus of the HV grid that interfaces the MV network, $P^{(\text{HV})}_{bt}$ and $Q^{(\text{HV})}_{bt}$ the active and reactive power flow at the substation transformer, and $P^{(\text{MV})}_{0t}$ and $Q^{(\text{MV})}_{0t}$ the active and reactive power at the slack bus of the MV grid. Substation transformers are included in the topology of the MV grids, therefore their losses do not appear here. The continuity on the apparent power flow reads as:
\begin{subequations} \label{eq:continuity}
\begin{align}
P^{(\text{HV})}_{bt} = P^{(\text{MV})}_{0t} \\
Q^{(\text{HV})}_{bt} = Q^{(\text{MV})}_{0t}.
\end{align}
for each $t$. For the voltage magnitude, we model the approximated operations of an OLTC by allowing the voltage at its secondary to vary within a certain range: 
\begin{align}
1/c \cdot v^{(\text{HV})}_{bt} \le v^{(\text{MV})}_{0t} \le c \cdot v^{(\text{HV})}_{bt}
\end{align}
where $c > 1$ is a design parameter. The expression above is linear, and it does not impact on convexity.
We assume that the impact of tap changes on the equivalent circuit model of the transformer (thus on the sensitivity coefficients of the grid) is negligible.
The optimal voltage magnitude at the MV slack is a decision variable of the problem.
%Since the admittance matrix of the network is unaltered, we assume that changing the position tap does alter the equivalent circuit model of the transformer}.
In this case, we do not impose any continuity constraint on the voltage phase angle between because the downstream network is single port and all its nodes consist of PQ injections that are invariant to phase angle. 
\end{subequations}

%{\color{red}In addition to the physical continuity constraints above, a limit on the power factor at the MV-to-HV interface is imposed as follows:
%\begin{align}  \label{eq:powerfactor}
%| Q^{(\text{MV})}_{0t}| \le  \text{tan}(\phi_{\text{min}}) \cdot |P^{(\text{MV})}_{0t}|,
%\end{align}
%where $\text{cos}(\phi_{\text{min}})$ is the minimum power factor limit.}
%\begin{align}
%- |P^{(\text{MV})}_{0t}| * \text{tan}(\phi_{\text{min}}) \le  Q^{(\text{MV})}_{0t} \le |P^{(\text{MV})}_{0t}| * \text{tan}(\phi_{\text{min}})
%\end{align}

% Metto questo nel case study description
%{\color{red} The minimum power factor  ($\text{cos}(\phi_{\text{min}})$) required in the simulations is 0.8, having considered a stressed system.}

\subsection{Conventional power plants}
Conventional power plants are modeled with constrained voltage-independent nodal active and reactive power injections, denoted by $P^{(G)}_b$ and $Q^{(G)}_b$, where $b$ is the node index.
%{\color{blue} In order to have a fair comparison between traditional and integrated regulation, conventional plants are assumed to have whatever regulating capacity is required by the system (regardless to power factor limitations), otherwise the traditional solution was too limited.}
They should respect the apparent power capacity $\bar{S}_{b}$ of the synchronous machine, the capability curve of the generator, and the minimum/maximum generating power capacity  $\underline{P}_{b}$ and $\bar{P}_{b}$ of the unit. As we focus on time constants related to secondary frequency regulation and we use input time series at 1 hour resolution, we do not model ramping rates, which normally refer to faster dynamics.

For all the nodes where generators are installed and for all time intervals $t=1, 2, \dots$, constraints read as follows:
\begin{subequations}\label{eq:convpp}
\begin{align}
    {P^{(G)}_{bt}}^2 + {Q^{(G)}_{bt}}^2 \le \bar{S}_{b}^2 \label{eq:convpp_1}\\
    |Q^{(G)}_{bt}| \le 0.8 \cdot P^{(G)}_{bt} \label{eq:convpp_2} \\
    \underline{P}_{b} \le P^{(G)}_{bt} \le \bar{P}_{b} \label{eq:convpp_3},
\end{align}
\end{subequations}
where 0.8 is the limit imposed by technical regulations \cite{CEI016}.

% --
%FABRI suggerirei qualcosa di meno specifico tipo:
% We assume that the power plant at the slack bus has enough capacity to achieve the energy balance in the system and we assign to it over-sized constraints. 
% -- 
% PAOLA intendevo esattamente il contrario: lo slack non regola niente perché ha profilo vincolato. Non specifichiamo che siam già lunghi.

% PAOLA Power plant located at slack node is included: even though the former equation cannot be applied (because active balance equation is imposed there), reactive reserve can be supplied by Generator 1.

% equazioni troppo specifiche:
%\begin{align}
%    \boldsymbol{P}_\text{min} \le \boldsymbol{P}_g + \Delta\boldsymbol{P}_g \le %\boldsymbol{P}_\text{max}. 
%\end{align}
%\begin{align}
%    -0.8 \left(\boldsymbol{P}_g + \Delta\boldsymbol{P}_g\right) \le \boldsymbol{Q}_g + %\Delta\boldsymbol{Q}_g \le 0.8 \left(\boldsymbol{P}_g + \Delta\boldsymbol{P}_g\right)
%\end{align}

\subsection{Demand and distributed generation} % ok
The power demand of the loads and the production of distributed PV generation at the various nodes of the grids are modeled with voltage-independent active and reactive power injections time series.
%{\color{red} power-controlled active and reactive injections, independent from voltage and admittance}.
For the case of the MV grids, time series are from measurements of real loads at the EPFL campus \cite{pignati2015real} of similar size as those considered. PV generation is simulated from measurements of the irradiance with the same modeling toolchain as in \cite{sossan2019solar}, that consists in transposing the irradiance, correcting it for the estimated cell temperature and air temperature, and scaling it for the average panel efficiency at Standard Test Conditions (STC).

%Since a limited number of time series for PV generation and loads is available, loads and PV generation for a large number of distributed resources are simulated by using measurements of similar days to avoid perfect correlation among nodal injections.
We assume that PV power plants are operated at unitary power factor, as in most current commercial configurations.

Day-ahead forecasts (used to linearize grid models and determine the unit commitment of the generators) are developed, for the demand, with a forecaster based on the $k$-nearest neighbor method as in \cite{7542590}, and, for a PV generation, starting from real forecast of the irradiance (provided by MeteoTest, Bern, CH) and processed with the same models described above.

For the injections of the lumped MV systems , we aggregate MV nodal injections until reaching the nominal demand of the respective primary substation.

\subsection{Energy storage systems} % ok
\paragraph{Modeling principle}
The injections of the ESSs are modeled as additive apparent power injections at each network node. In other words, each nodal injection at each time interval is the algebraic sum of the original apparent power injection (given by the demand or distributed generation time series) and two free variables (one for active power, another for reactive) which model the potential contribution of the battery. In this way, at each time interval, the free variables are "modulated" to allow the optimal power flow for solving grid congestions or for providing reserve according to the cost function. The activation of a free variable (thus, battery injection) at a certain node denotes that a ESS should be installed at that location.  This intuition explains how the siting principle works. The battery energy storage capacity and the rating of the associated power converter are derived from the evolution over time of the battery injection as now explained.

\paragraph{Apparent power rating of the power converter} % ok
Let $P^{(B)}_{tb}, Q^{(B)}_{tb}$ denote the battery injection at node $b$. Then the apparent power rating $S_{\text{nom}, b}$ of the power converter at node $b$ is the maximum apparent power observed over time (maximum of norms, convex):
\begin{align}
S_{\text{nom}, b} = \underset{t}{\text{max}} \left\{ \sqrt{{P^{(B)}_{tb}}^2 +  {Q^{(B)}_{tb}}^2 },\  t = 1,2,\dots \right\}, \label{eq:powerrating}
\end{align}
We assume that the power capability of the power converter does not depend on the grid voltage.

\paragraph{Energy capacity} % ok
For evaluating the required energy capacity, we first introduce the notion of battery state-of-energy (SOE). For the moment, we assume a lossless battery (the extension to a non-ideal battery is straightforward and implemented as described in the next paragraph), so that the SOE at time $t$ is the discrete integral over time of the battery injection (assuming zero initial state-of-charge):
\begin{align}
\text{SOE}_{tb} = T_s \sum_{\tau=1}^{t} P^{(B)}_{\tau b}, \label{eq:stateofenergy}
\end{align}
where $T_s$ is the sampling time. The value of the series $\text{SOE}_{tb}$ can be regarded to as the energy capacity required to accomplish the active power trajectory $P_{1b}, P_{2b}, \dots, P_{tb}$. One could simply estimate the required energy capacity by evaluating the final value of the time series. However, in this way, the energy capacity would depend on the length of the optimization horizon, thus invalidating this sizing principle. For instance, batteries persistently charging or discharging to compensate recurrent daily over-voltages would have a monotonic SOE pattern (eventually unbounded). To avoid patterns of this kind, we enforce that the SOE at the end of each day is the same as the starting value. In this context,
%Therefore, we emerge the operational consideration that batteries can be recharged or discharged from time to time, thus allowing for adjusting their state-of-charge to a suitable level. This process can be, for instance, achieved as described in \cite{7542590}, where the charging/discharging demand of batteries with depleted flexibility is embedded in the day-ahead forecasts of the aggregated demand and committed in a day-ahead market. This process is reasonable and in-line with power system operational practices, which typically require to schedule the operation of power plants (especially the least flexible) with a certain clearing on and for a certain horizon.
the required energy capacity for each day can be estimated by evaluating expression \eqref{eq:stateofenergy} in blocks of 1-day duration, and for each taking the difference between the maximum and the minimum values. For the formal definition, we proceed with assuming a scheduling horizon of 1 day for the sake of clarify. Any other time interval can be equally accommodated by adapting the following definitions.
We first introduce the sets $\Gamma_{\texttt{day0}}, \Gamma_{\texttt{day 1}}, \dots$; each contains all the time indexes that belong to the respective day.
The battery energy capacity required for one day of operation
\begin{align}\label{eq:Eb_day}
\begin{aligned}
E_{b}(\texttt{day0}) =
& \underset{t}{\text{max}} \left\{ \text{SOE}_t, \forall t \in \Gamma_{\texttt{day0}} \right\} - \\
 + &\underset{t}{\text{min}} \left\{ \text{SOE}_t \forall t \in \Gamma_{\texttt{day0}} \right\},
\end{aligned} 
\end{align}
and similarly for the other days. The final energy storage capacity required at node $b$ is modeled as the maximum value over all days:
\begin{align}
E_{\text{nom}, b}  &= \underset{\text{days}}{\text{max}} \left\{
E_{b}(\text{day 0}), E_{b}(\text{day 1}), \dots \right\}. \label{eq:energycapacity}
\end{align}
By forcing the SOE at the end of each day to return to its initial value, we ensure that the problem does not take advantage of the starting energy stock.
%, which does not affect the results thank to \eqref{eq:Eb_day}.}
%NOTE: here we are assume we can always charge the battery without violating grid constraints.

\begin{figure}
    \centering
    \includegraphics[width=0.45\columnwidth]{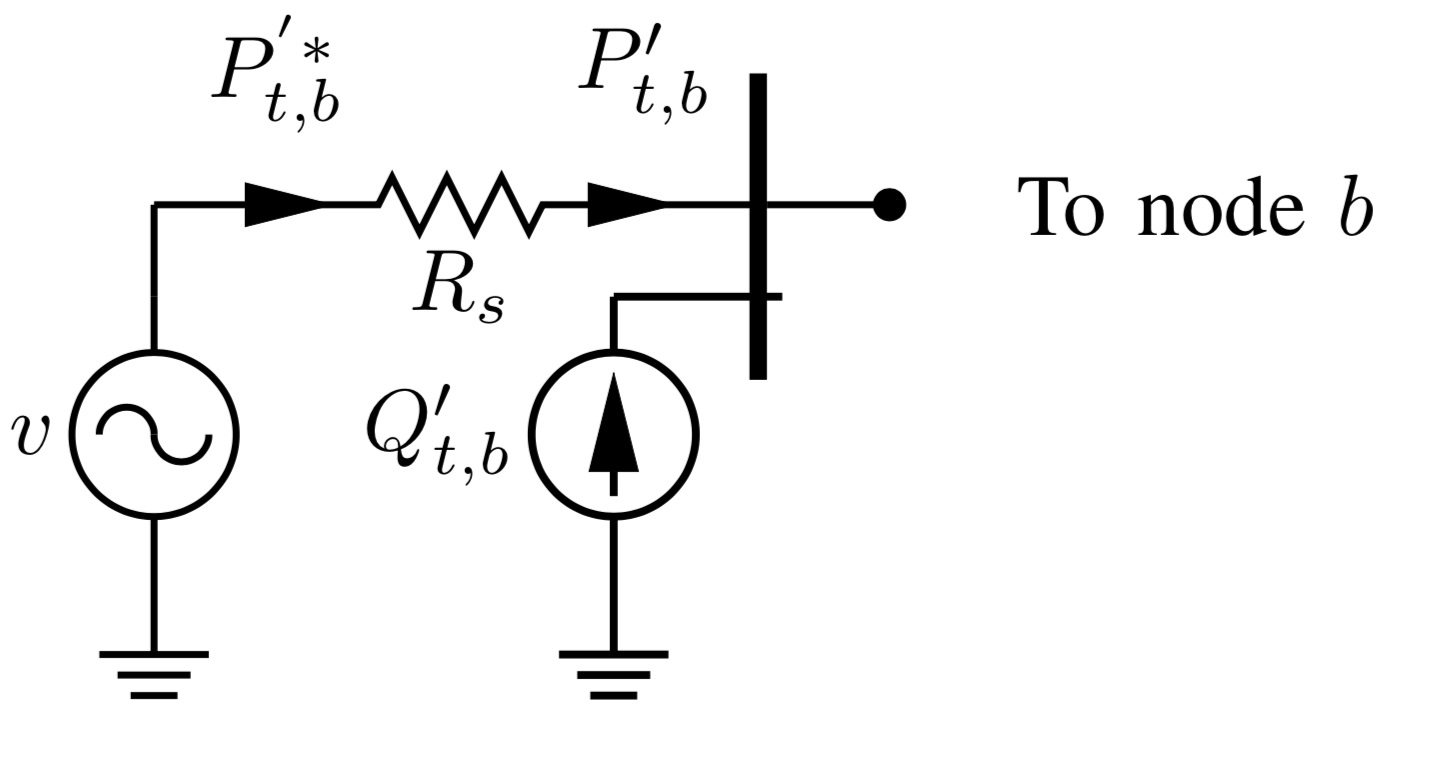}
    \caption{Equivalent circuit model of the battery energy storage system. $P'_{t,b}, Q'_{t,b}$ are the active and reactive power injections of the battery. $P'_{t,b}$ is the 'lossy' active power injection that is used to drive the model of state-of-energy of the battery.}
    \label{fig:battery}
\end{figure}

\paragraph{Model properties and approximation}
Expression \eqref{eq:stateofenergy} did not account for energy conversion losses. To model them, we use the approach proposed in \cite{stai_2017} that relies on an approximated Thevenin equivalent circuit of whole battery's power conversion toolchain that is added in series to node of the grid that hosts the battery, thus achieving a seamless integration of the notion of charging/discharging efficiency in the load flow problem. Fig.~\ref{fig:battery} shows the equivalent circuit model: $P'_{t,b}$ and $Q'_{t,b}$ are the active and reactive power injections of the battery as seen from the grid, whereas $P^{*'}_{t,b}$ is the lossy battery power output that feeds the battery state-of-energy model in \eqref{eq:stateofenergy}. The new controllable variables of the model are $P^{*'}_{t,b}$ and $Q^{'}_{t,b}$, whereas the injection $P'_{t,b}$ is calculated by the grid model. The additional node is a modelling abstraction to represent conversion losses, and no grid voltage and line current constraints are added to that.

\subsection{Decision problem}
\subsubsection{Model of the power balance in the system}
During real-time operations, stochastic realizations of the demand and distributed PV deviate from forecasted profiles. To maintain the balance, injections of the power plants and of ESSs will be activated according to the cost function described next. Let $\tilde{P}^{(\text{HV})}_{0t}$ the real power flow at the slack bus of the HV network at time $t$ determined by the day-ahead unit commitment\footnote{Computed by an optimal power flow fed with with day-ahead point predictions of the stocastic realizations.}.
The active power balance of the system is modeled by enforcing that the realization of the active power flow at the slack bus $P^{(\text{HV})}_{0t}$ matches with the respective day-ahead flow $\tilde{P}^{(\text{HV})}_{0t}$: 
\begin{align}
P^{(\text{HV})}_{0t}(\cdot) = \tilde{P}^{(\text{HV})}_{0t} \label{eq:powerbalance},
\end{align}
where $P^{(\text{HV})}_{0t}(\cdot)$ is a function of all controllable injections (ESSs and conventional generators) and stochastic realizations and it is computed with \eqref{eq:gridmodel2}. By imposing the slack power to equal the day-ahead commitment with \eqref{eq:powerbalance}, we force all the other units, including ESSs in distribution grids, to provide balancing power.

\subsubsection{Cost of operations}
Let $c_{bt}$ be the (symmetric) cost of activating a unit of power reserve for the time interval $t$ and for the power plant at bus $b$. $\mathcal{G}$ is the set of bus indexes that interface conventional power plants. We assume that reserve is provided at the day-ahead spot price\footnote{For up regulation this can be a very conservative estimation, see for instance \cite{skytte1999regulating}.}. The total cost of providing regulation from conventional power plants is the sum over time and over all the plants:
\begin{align}
    J^{(G)}\left(\boldsymbol{\Delta P}^{(G)}\right) = \sum_{t \in \mathcal{T}, b \in \mathcal{G}} c_{bt} \Delta P^{(G)}_{bt},
\end{align}
where $\boldsymbol{P}^{(G)}$ collects all the power injections of conventional generators. 
The costs related to batteries are the capital investments required to manufacture them. In doing so, we assume that batteries can be essentially recharged (and discharged) for free, for instance, by taking advantage of low prices in the day-ahead spot market (viceversa).  Fixed installation costs are disregarded. Let $\mathcal{B}$ be the set of bus indexes of both HV and MV grids candidate for hosting a ESS\footnote{Not all the nodes might be available to host an ESS due to, for instance, land use constraints.}, and $c_{P}$ and $c_{E}$ be the unitary cost of apparent power rating (\euro/MVA) and energy storage capacity (\euro/MWh). 
With reference to models \eqref{eq:powerrating} and \eqref{eq:energycapacity}, the cost of installing ESSs is:
\begin{align}
    \begin{aligned}
        J^{(B)}\left( \boldsymbol{P^{(B)}}, \boldsymbol{Q^{(B)}} \right) = \\
        \sum_{b \in \mathcal{B}} 
        \left(
        c_{P}S_{\text{nom}, b}\left(\boldsymbol{P^{(B)}_b}, \boldsymbol{Q^{(B)}_b}\right) + c_{E}E_{\text{nom}, b}\left(\boldsymbol{P^{(B)}_b}\right)
        \right),
    \end{aligned}
\end{align}
where variables $\boldsymbol{P^{(B)}}, \boldsymbol{Q^{(B)}}$ collect all ESSs' active and reactive power injections.
The total system cost is the sum of the last two:
\begin{align}
    J(\cdot) = J^{(G)}\left(\boldsymbol{\Delta P}^{(G)}\right) + J^{(B)}\left( \boldsymbol{P^{(B)}}, \boldsymbol{Q^{(B)}} \right)\label{eq:cost}
\end{align}
where the arguments of $J$ are omitted for brevity.

\subsubsection{Complete formulation}
The decision problem consists of minimizing costs \eqref{eq:cost} subject to the power balance constraint in \eqref{eq:powerbalance}, conventional power plants constraints \eqref{eq:convpp}, ESSs' power rating \eqref{eq:powerrating} and energy storage requirements \eqref{eq:energycapacity}, grid models \eqref{eq:gridmodel} for both the HV and MV grids, continuity constraints \eqref{eq:continuity}%and power factor at interface \eqref{eq:powerfactor}
, grid constraints on statutory voltage limits and line ampacities. We model multiple daily scenarios of the realizations (by stacking them along the time dimensions) so to compute storage deployment guidelines that are valid for multiple stochastic outcome.
The formal formulation is omitted due to lack of space.

%\clearpage

%That is convex since it is the positive sum of the point-wise maximums of affine functions. Energy can even become negative during the day, so that the starting point does not influence the result (only the difference is taken into account for sizing). In each daily scenario, the final State-of-Energy has been constrained to initial value, in order not to take advantage of ''starting'' energy.
%In the cost function, equivalent daily ESS CAPEX cost is used, amortized on a 20-year lifetime.

\section{Case study}
We provide a proof-of-concept of the proposed formulation by numerical simulations considering the system shown in Fig.~1. Grid topologies for the HV and MV systems are according to the specifications of the 9-bus IEEE test system and CIGRE' benchmark system for MV grids, respectively. The nominal demand and PV capacity at each node are reported in \ref{tab:nodenominaldemand}.
We model power demand as described in Section III and with nominal nodal values according to the specifications of the benchmark systems. Instead, the PV installed capacity is chosen to create mild reverse power flow conditions during peak production hours to reflect conditions of future distribution grids with pervasive distributed generation. Equivalent loads A and B in Fig. \ref{fig:Case_Study} are modelled also starting from real measurements: the former refers to MV a grid with large presence of PV generation, the second to a grid with loads only.

\begin{table}[h]
\centering
\caption{Nominal demand, power factor ($pf$) and installed PV capacity of traditional generation per node}
\label{tab:nodenominaldemand}
\scriptsize{\def\arraystretch{1.1}
\begin{tabular}{ | c | c | c | c | c | c | c | }
\hline
\multirow{3}{*}{Node} & \multicolumn{3}{c|}{MV network} & \multicolumn{3}{c|}{HV network} \\
   \cline{2-7}
   & \multicolumn{2}{c|}{Residential demand} & PV & \multicolumn{2}{c|}{Nominal demand}  & Generation   \\
   \cline{2-7}
 & MVA   & $pf$   & MWp 	& MW  & Mvar & MVA-MWp \\
 \hline
1 & 15.3 & 0.9 & 5.7 	  &  0     & --   & 250 \\
2 &   0  & --   & 5.0 	  &  0     & --   & 300 \\
3 & 0.28 & 0.9 & 4.4 	  &  0     & --   & 270 \\
4 & 0.44 & 0.9 & 2.2 	  &  0     & --   & 0   \\
5 & 0.75 & 0.9 & 1.3 	  & 35     & 30   & 40 \\ 
6 & 0.56 & 0.9 & 0.1 	  &  0     & --   & 0   \\
7 & 0 	 & --   & --   	  & 100    & 50   & 0   \\
8 & 0.6  & 0.9 & --  	  &  0     & --   & 0   \\
9 & 0 	 &  --  & 1.1    & (34.3) & (31) & (37.8) \\
10 & 0.49 & 0.9 & 1.6   &       &      &    \\
11 & 0.34 & 0.9 & 5.7   &       &      &    \\
12 & 15.3 & 0.9 & 5.7   &       &      &    \\
13 & 0 	 & -- 	 & --     &       &      &    \\
14 & 0.22 & 0.9 & 5.0   &       &      &   \\
\hline
\bf Total & 34.28 &  & 37.8 & 169.3 & 111 & 820-77.8 \\
\hline
\end{tabular}
}
\end{table}

To model PV generation and demand, we consider four typical days of PV generation time series, one per season. For the demand, we consider the same daily profile for all four days because the considered measurement data set does not include thermal loads and, thus, does not exhibit significant seasonal patterns, that are, instead, dominated by PV generation.
For each day, we consider 5 possible scenarios of the stochastic PV generation and demand realization, for a total of 20 scenarios. At the current stage, the number of scenarios is chosen so as to have tractable computational times. An higher number of scenarios with guaranteed robust performance, as in \cite{campi2008exact}, will be considered in future works.
Scenarios are derived as follows considering real measurements and forecast for Lausanne (CH) from 2018. For each season, we first group similar PV generation forecasts with the $k$-means algorithm, where k is estimated with the silhouette analysis. The cluster with the largest number of elements is chosen as the most representative for the current season and retained for the next analysis. For each forecast series in the retained cluster, we select the associated realization of PV production (from measurements) and consider them as a possible scenario. These scenarios are reduced to the final number by re-applying the $k$-means algorithm with $k=5$.

We consider a constant cost of the power reserve from conventional generation of 50 \euro/MWh. This value is chosen because it is near the average spot price in Europe (see, e.g., \cite{bang2012existing}) and a good proxy for the regulation price \cite{skytte1999regulating}. Reactive compensation is "for free" and subject to the capability curves of the generators.
The operational constraints enforced in the decision problem refer to voltage limits for all HV and MV nodes in the range 0.95-1.05 pu, power factor at the primary substation equal to or larger than 0.8, tap changer position at the middle plus/minus 10\%.
%The constraint on the minimum power factor at the MV-to-HV interface is 0.8.
%Line current constraints, when not active (verified with a load flow after the problem was solved), were omitted to improve the computational time and include more scenarios. %It is verified ex post that limits are never exceeded.

The costs of batteries costs are 280 \euro/kWh for energy capacity \cite{ESSprices}, and 80 \euro/kVA for power converter rating.

\section{Results}
We first show the output of the proposed method, that refers to the locations, and apparent power rating and energy capacity requirements of the battery systems to deploy. In a second illustrative analysis, we show that by forcing batteries' injections to zero (so as to activate reserve from conventional power plants only), the problem is feasible and converges at a similar cost only when constraints are relaxed, thus demonstrating that the use of battery energy storage systems improves grid control performance while subject to similar implementation costs.

\subsection{Vertical and horizontal siting and sizing}
\begin{figure}[!h]
\centering
\includegraphics[width=0.90\columnwidth]{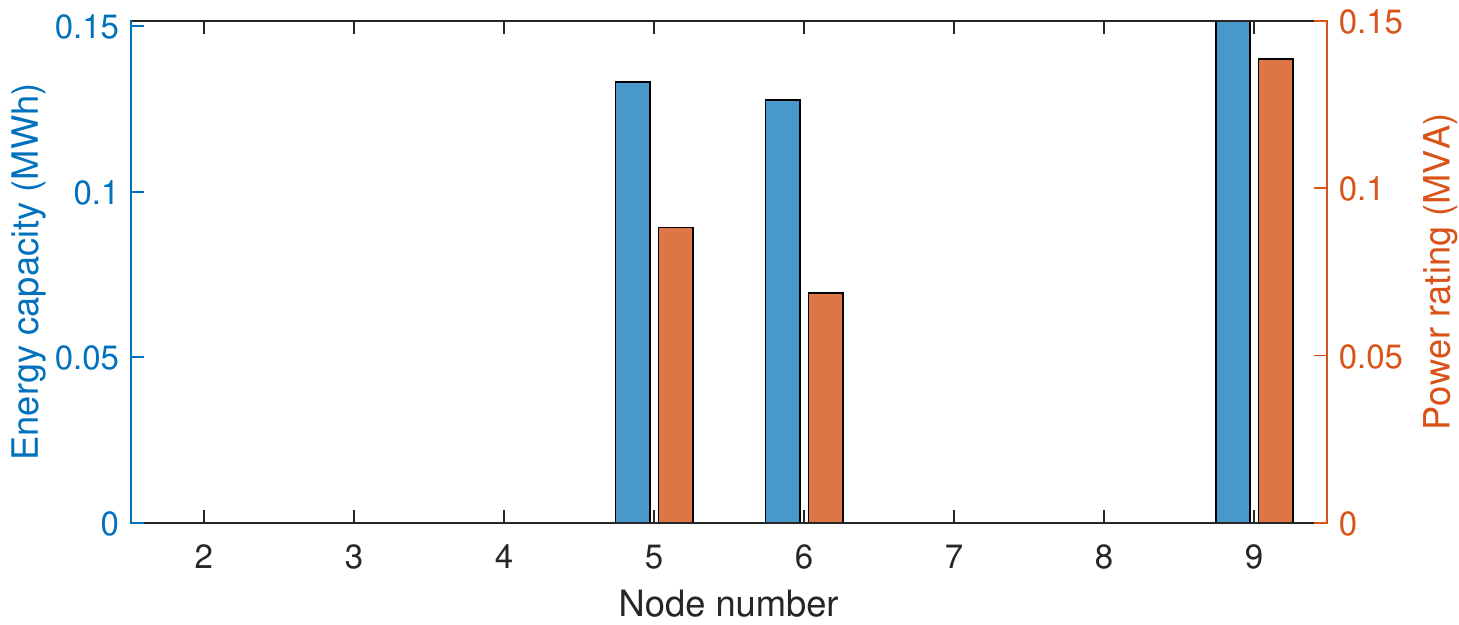}
\caption{Sizing requirements as a function of the nodal locations in HV grid.}
\label{fig:Installed_BESS_HV}
\end{figure}

\begin{figure}[!h]
\centering
\includegraphics[width=0.90\columnwidth]{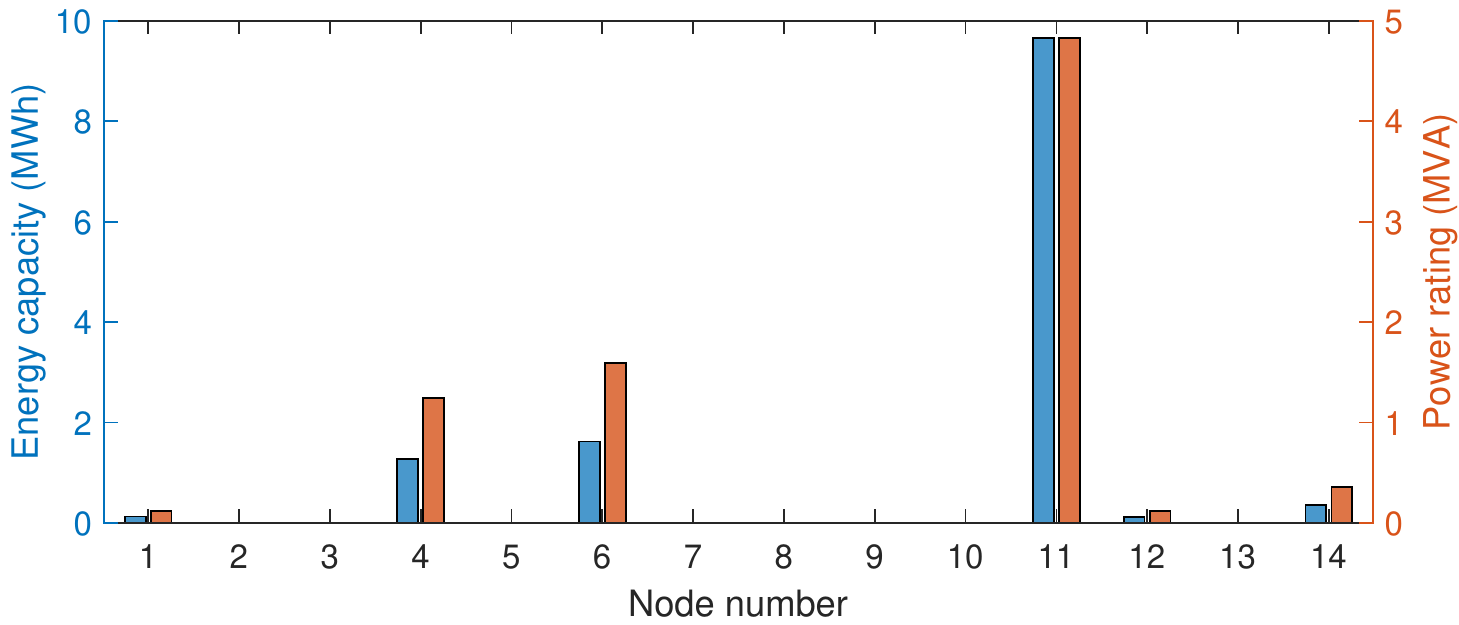}
\caption{Sizing requirements as a function of the nodal locations in MV grid.}
\label{fig:Installed_BESS_MV}
\end{figure}

The siting and sizing results for the HV and MV grids are shown in Figures \ref{fig:Installed_BESS_HV} and \ref{fig:Installed_BESS_MV}, respectively. In the HV grid, three ESSs are installed: at nodes 5 (together with aggregated load A), 6 (where a generator is also connected, as visible in Fig.~\ref{fig:Case_Study}), and 9. The C-rate (i.e., active power rating over energy capacity) of the installed ESSs is between 1/2 and 1, which is well within the technical capability range of commercially available Lithium-ion ESSs. The relatively low C-rate of the proposed application denote an energy-intensive use of ESSs and can be therefore couple well with power-intensive services, such as primary frequency control. In the MV grid, the largest ESS is installed at the end of the first feeder, at node 11 (almost  1/2C). Smaller devices are installed at nodes 6, 4, 14, 1, and 12, all with C-rates slightly smaller than 1.

\begin{table}[!h]
\centering
\caption{Optimal ESS sizes and cost comparison against generators' reserve}
\label{Tab:results_cost}
{\scriptsize \def\arraystretch{1.1}
\begin{tabular}{|l|l|c|c|}
\hline
\multicolumn{2}{|l|}{\textbf{Complete optimization results}}  & ESS size  & Cost  \\ \hline
\multirow{2}{*}{HV}            & Energy capacity            & 0.41 MWh-0.30MW  & 0.12 M\euro        \\ %\cline{2-4}
    & Power rating               & 0.46 MVA   & 0.04  M\euro       \\ \hline
\multirow{2}{*}{MV}            & Energy capacity            & 13.16 MWh-8.25MW  & 3.68 M\euro       \\ %\cline{2-4} 
                               & Power rating               & 19.30 MVA   & 1.54  M\euro        \\ \hline
\multicolumn{2}{|l|}{Generators active reserve}  & 157.05 GWh  & 7.85 M\euro        \\ \hline 
\multicolumn{2}{|l|}{Total cost}    &   -       & 13.23 M\euro      \\ \hline \hline
\multicolumn{2}{|l|}{\textbf{Comparison with traditional reserve}}  & Total energy  & Cost  \\ \hline
\multicolumn{2}{|l|}{Generators active reserve} %& 20.09 TWh  & 2008.92 M\euro \\ \hline
& 303.67 GWh  & 15.18 M\euro \\ \hline
\hline
\multicolumn{2}{|l|}{\textbf{Reactive injections required}}  & Mean       & Max          \\ \hline
\multicolumn{2}{|l|}{HV ESS reactive injections}           & 0.05 MVAR  & 0.18 MVAR   \\ %\hline
\multicolumn{2}{|l|}{MV ESS reactive injections}           & 1.50 MVAR  & 6.94 MVAR   \\ %\hline
\multicolumn{2}{|l|}{HV gen. reactive injections}           & 5.25 MVAR  & 40.08 MVAR   \\ \hline %\hline
\multicolumn{2}{|l|}{Only with traditional reserve} & 7.78 MVAR & 45.09 MVAR   \\ \hline 
\end{tabular}}
\end{table}

Table \ref{Tab:results_cost} reports the total ESSs requirements and costs. In order to compare the capital cost for the ESSs installation and the operational cost for reserve provision from conventional generation, we project the costs of reserve over the life-span of the energy storage facilities, that we assume of 20 years. At this stage we consider calendar ageing only, that for certain battery technologies, such as lithium-titanate, is dominant over the power cycling aging.
% TODO revise this part according to new results
%Required mean and maximum reactive injections are non-uniformly spread in the networks, especially at MV level, where the optimal power rating of converters can be significantly higher than the actual ESSs size (especially at node 14, end of the second feeder) due to the absence of other reactive sources.
%Most of the available devices, indeed, are often required to work close to their capability limits.
%It is worth noting that for voltage regulation, both HV and MV slack regulations (including tap changer) are largely leveraged by the optimization.

\begin{figure}[!h]
\centering
\footnotesize
\input{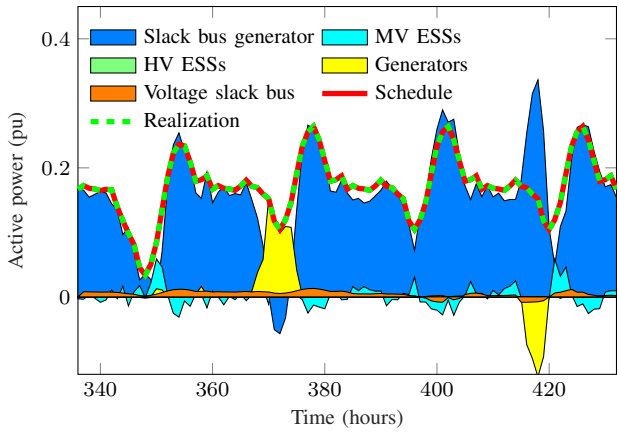}
\caption{Compensation of the imbalances by the various reserve providers.}
\label{fig:Power_balance}
\end{figure}

Fig.~\ref{fig:Power_balance} shows how the different reserve providers contribute to achieving the power balance and the day-ahead schedule in a certain time period: the contribution of the ESSs in MV grids (cyan area) complements the injections of the generators at the slack bus (blue) and at the other nodes (yellow) so to ensure that the total generation (dashed green) follows the schedule (red). As visible in Fig. \ref{fig:Power_balance}, also variations of the HV slack voltage (orange) for voltage regulation determines variations of the power balance due to different levels of power losses in the lines.

\begin{figure}[!h]
\centering
\footnotesize
\input{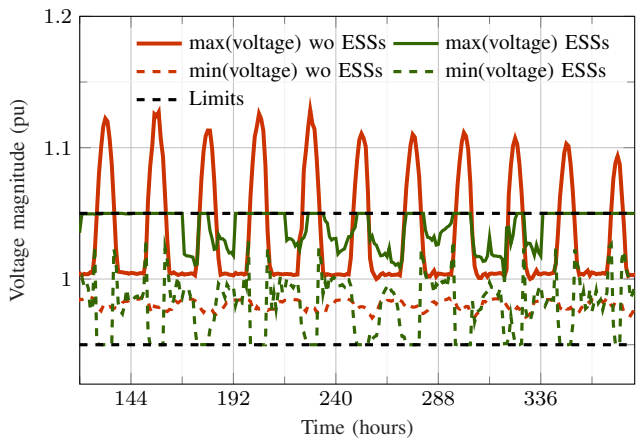}
\caption{Evolution in time of the maximum and minimum voltage magnitudes over all the grid nodes before and after ESSs actions for a selected period. Voltage limits are 1 pu$\pm$5\%.}
\label{fig:Voltage_profiles}
\end{figure}

\begin{figure}[!h]
\centering
\footnotesize
\input{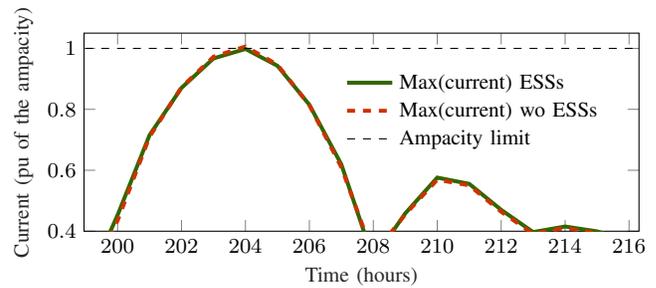}
\caption{Evolution in time of the maximum and minimum line currents over all the lines before and after ESSs actions for a given period where a mild violation occurs. Values are reported in per unit of the line's ampacity limit.}
\label{fig:currents}
\end{figure}

Figures \ref{fig:Voltage_profiles} and \ref{fig:currents} shows the voltage magnitudes and line currents (the seconds expressed in per unit of the respective ampacity limit) before and after the control action of the batteries for a period of interest where violations occur. The considered case study is dominated by voltage violations, whereas Line ampacity violations are less frequent and mild. As visible in Fig. \ref{fig:Voltage_profiles}, voltage levels are often considerably above the 1.05~pu limit due to PV generation. The action of distributed storage is able to restore voltage levels within the prescribed limits. Current violations are small (up to 1\% recorded during the 5 PV-peak-production hours of the summer period in the MV grid) and are also corrected by distributed ESSs.
%Fig. \ref{fig:Voltage_profiles} shows the voltage magnitudes of all nodes over time with (green) and without (blue) ESSs, and how they compare to the selected voltage statutory voltage limits. It denotes that the batteries actions improve voltage levels, and they achieve constraints satisfaction.

\subsection{Comparison with conventional generation only}
Due to excess PV generation in the MV networks, the problem without ESSs is not feasible under the tight constraints of the previous case study. Thus, we allow larger constraints on the voltage magnitude (from $\pm$5\% to $\pm$7\%). Power plants are assumed to have enough capacity to provide all the reserve required by the system.
The costs for reserve procurement projected on 20 years is \euro~15.18~million, thus higher than the former case (i.e., \euro~13.23~million, Table II). 
%We can conclude that the use of ESSs starts to be economically justifiable above the considered price of regulating power (\euro~90 per MWh). At the same time
It is important to acknowledge that ESSs achieved to implement voltage control in the MV network compared to the case with conventional generation only; if this service is remunerated in the future, it will contribute to shortening the payback time of ESSs. The average reactive injections supplied by conventional generators when ESSs are not available is 7.78 MVAR, with a maximum of 45.09 MVAR.

\section{Conclusions}
We proposed a modeling framework to determine the optimal location, energy capacity and power rating of distributed battery energy storage systems at multiple voltage levels for local grid control (voltage regulation and congestion management) and reserve provision to the transmission system operator. The decision model relies on an optimal power flow problem with a representation of the grid constraints of MV and HV grids and the regulation capacity of conventional generation units connected at the HV level. We model grid operations at a one-hour resolution, where stochastic realizations of the demand and distributed generation determine power imbalances from a day-ahead schedule. The imbalances are compensated by changing the set-point of conventional power plants (i.e., conventional reserves) and, possibly, by injections of the batteries. By assigning an operational cost to conventional reserves and a capital cost to batteries’ power rating and energy capacities, we derive the technical-economical optimum for storage systems deployment. Batteries injections are also activated to enforce grid constraints (in case, for example, of excess distributed PV generation in distribution grids) and, thus, provide both regulating power and perform grid control. Because we consider the needs of both distribution and transmission system operators, we refer to this formulation as vertical and horizontal planning of energy storage systems, as opposed to horizontal planning that includes a single voltage level only.  We use linearized models of the HV and MV grids and retain fundamental modeling aspects of the power system (transmission losses, effects of reactive power, OLTC at the MV/HV interface, non-unitary efficiency of battery energy storage systems) with a tractable convex formulation of the decision problem. Demand and distributed generation are modeled with a bottom-up approach with real measurements and forecasted with methods from the literature. We model the stochastic outcomes of the loads and distributed generation with scenarios.

We test the performance of the planning algorithm by simulations on a joint power system that includes the IEEE 9-bus system and the CIGRE’ benchmark model for MV systems. Results show that the proposed method determines energy storage deployment plans that meet the required specifications. A comparison against the case without storage denotes that the capital costs for ESSs deployment are competitive compared to accumulated operational costs for reserves and that ESSs successfully achieves voltage regulation in grids with large amounts of PV generation. 

Vertical and horizontal planning determines an optimized strategy for the deployment of energy storage systems to serve the needs both the needs of transmission and distribution grids and is, therefore, a useful resource for system planners. Moreover, it can also be adopted by policy-makers to design specific policies encouraging the deployment of energy storage systems in distribution networks. Future developments are in the direction of developing a distributed formulation of the load problem to improve the scalability of the algorithm and extend its application to large networks. In addition to identifying the technical-economical optimum for storage deployment, this work also opens to the possibility of designing optimal policies and incentives to foster the adoption of storage accounting for its inherent multi-service nature. 

%{\color{red} We may add a scalability analysis with respect to HV and MV nodes and lines, to show applicability to real asses?}

%Future developments will include installation costs, full constraint implementation, sensitivity to market reserve prices, test possible impact of ramping constraints, fasten computation with current constraints...

%{\footnotesize
%\section*{References}
%\noindent [1] Nick et al. Optimal allocation of dispersed energy storage systems in active distribution networks for energy balance and grid support. IEEE Transactions on Power Systems. 2014.\\
%\noindent [2] Pandžić et al. Near-optimal method for siting and sizing of distributed storage in a transmission network. IEEE Transactions on Power Systems. 2014.
%}

\bibliographystyle{IEEEtran}
\bibliography{Biblio.bib}

\begin{thebibliography}{10}
\providecommand{\url}[1]{#1}
\csname url@samestyle\endcsname
\providecommand{\newblock}{\relax}
\providecommand{\bibinfo}[2]{#2}
\providecommand{\BIBentrySTDinterwordspacing}{\spaceskip=0pt\relax}
\providecommand{\BIBentryALTinterwordstretchfactor}{4}
\providecommand{\BIBentryALTinterwordspacing}{\spaceskip=\fontdimen2\font plus
\BIBentryALTinterwordstretchfactor\fontdimen3\font minus
  \fontdimen4\font\relax}
\providecommand{\BIBforeignlanguage}[2]{{%
\expandafter\ifx\csname l@#1\endcsname\relax
\typeout{** WARNING: IEEEtran.bst: No hyphenation pattern has been}%
\typeout{** loaded for the language `#1'. Using the pattern for}%
\typeout{** the default language instead.}%
\else
\language=\csname l@#1\endcsname
\fi
#2}}
\providecommand{\BIBdecl}{\relax}
\BIBdecl

\bibitem{koller2015review}
M.~Koller, T.~Borsche, A.~Ulbig, and G.~Andersson, ``Review of grid
  applications with the zurich 1 mw battery energy storage system,''
  \emph{Electric Power Systems Research}, vol. 120, pp. 128--135, 2015.

\bibitem{OULDAMROUCHE201620914}
\BIBentryALTinterwordspacing
S.~{Ould Amrouche}, D.~Rekioua, T.~Rekioua, and S.~Bacha, ``Overview of energy
  storage in renewable energy systems,'' \emph{International Journal of
  Hydrogen Energy}, vol.~41, no.~45, pp. 20\,914 -- 20\,927, 2016. [Online].
  Available:
  \url{http://www.sciencedirect.com/science/article/pii/S0360319916309478}
\BIBentrySTDinterwordspacing

\bibitem{7542590}
F.~Sossan, E.~Namor, R.~Cherkaoui, and M.~Paolone, ``Achieving the
  dispatchability of distribution feeders through prosumers data driven
  forecasting and model predictive control of electrochemical storage,''
  \emph{IEEE Transactions on Sustainable Energy}, vol.~7, no.~4, Oct 2016.

\bibitem{7429781}
E.~{Serban}, M.~{Ordonez}, and C.~{Pondiche}, ``Voltage and frequency grid
  support strategies beyond standards,'' \emph{IEEE Transactions on Power
  Electronics}, vol.~32, no.~1, pp. 298--309, Jan 2017.

\bibitem{8464265}
F.~{Conte}, F.~{D’Agostino}, P.~{Pongiglione}, M.~{Saviozzi}, and
  F.~{Silvestro}, ``Mixed-integer algorithm for optimal dispatch of integrated
  pv-storage systems,'' \emph{IEEE Transactions on Industry Applications},
  vol.~55, no.~1, pp. 238--247, Jan 2019.

\bibitem{STAFFELL2016212}
\BIBentryALTinterwordspacing
I.~Staffell and M.~Rustomji, ``Maximising the value of electricity storage,''
  \emph{Journal of Energy Storage}, vol.~8, pp. 212 -- 225, 2016. [Online].
  Available:
  \url{http://www.sciencedirect.com/science/article/pii/S2352152X1630113X}
\BIBentrySTDinterwordspacing

\bibitem{7422904}
M.~{Zidar}, P.~S. {Georgilakis}, N.~D. {Hatziargyriou}, T.~{Capuder}, and
  D.~{Škrlec}, ``Review of energy storage allocation in power distribution
  networks: applications, methods and future research,'' \emph{IET Generation,
  Transmission Distribution}, vol.~10, no.~3, pp. 645--652, 2016.

\bibitem{wu2015energy}
D.~Wu, C.~Jin, P.~Balducci, and M.~Kintner-Meyer, ``An energy storage
  assessment: Using optimal control strategies to capture multiple services,''
  in \emph{2015 IEEE Power \& Energy Society General Meeting}.\hskip 1em plus
  0.5em minus 0.4em\relax IEEE, 2015, pp. 1--5.

\bibitem{7932132}
M.~{Kazemi}, H.~{Zareipour}, N.~{Amjady}, W.~D. {Rosehart}, and M.~{Ehsan},
  ``Operation scheduling of battery storage systems in joint energy and
  ancillary services markets,'' \emph{IEEE Transactions on Sustainable Energy},
  vol.~8, no.~4, pp. 1726--1735, 2017.

\bibitem{Namor_tsg_2018}
E.~{Namor}, F.~{Sossan}, R.~{Cherkaoui}, and M.~{Paolone}, ``Control of battery
  storage systems for the simultaneous provision of multiple services,''
  \emph{IEEE Transactions on Smart Grid}, vol.~10, no.~3, pp. 2799--2808, May
  2019.

\bibitem{zidar2016review}
M.~Zidar, P.~S. Georgilakis, N.~D. Hatziargyriou, T.~Capuder, and
  D.~{\v{S}}krlec, ``Review of energy storage allocation in power distribution
  networks: applications, methods and future research,'' \emph{IET Generation,
  Transmission \& Distribution}, vol.~10, no.~3, pp. 645--652, 2016.

\bibitem{Mostafa_ESSplan}
M.~{Nick}, R.~{Cherkaoui}, and M.~{Paolone}, ``Optimal planning of distributed
  energy storage systems in active distribution networks embedding grid
  reconfiguration,'' \emph{IEEE Transactions on Power Systems}, vol.~33, no.~2,
  pp. 1577--1590, March 2018.

\bibitem{Pandzic}
H.~{Pandžić}, Y.~{Wang}, T.~{Qiu}, Y.~{Dvorkin}, and D.~S. {Kirschen},
  ``Near-optimal method for siting and sizing of distributed storage in a
  transmission network,'' \emph{IEEE Transactions on Power Systems}, vol.~30,
  no.~5, pp. 2288--2300, Sep. 2015.

\bibitem{CELLI2018154}
\BIBentryALTinterwordspacing
G.~Celli, F.~Pilo, G.~Pisano, and G.~G. Soma, ``Distribution energy storage
  investment prioritization with a real coded multi-objective genetic
  algorithm,'' \emph{Electric Power Systems Research}, vol. 163, pp. 154 --
  163, 2018. [Online]. Available:
  \url{http://www.sciencedirect.com/science/article/pii/S0378779618301810}
\BIBentrySTDinterwordspacing

\bibitem{fortenbacher2016optimal}
P.~Fortenbacher, M.~Zellner, and G.~Andersson, ``Optimal sizing and placement
  of distributed storage in low voltage networks,'' in \emph{2016 Power Systems
  Computation Conference (PSCC)}.\hskip 1em plus 0.5em minus 0.4em\relax IEEE,
  2016, pp. 1--7.

\bibitem{8918388}
H.~{Abdeltawab} and Y.~A.~I. {Mohamed}, ``Mobile energy storage sizing and
  allocation for multi-services in power distribution systems,'' \emph{IEEE
  Access}, vol.~7, pp. 176\,613--176\,623, 2019.

\bibitem{Powertech}
{F. D'Agostino, S. Massucco, P. Pongiglione, M. Saviozzi, F. Silvestro},
  ``Optimal der regulation and storage allocation in distribution networks:
  Volt/var optimization and congestion relief,'' \emph{IEEE Milan PowerTech},
  June 2019.

\bibitem{8302948}
A.~{Hassan} and Y.~{Dvorkin}, ``Energy storage siting and sizing in coordinated
  distribution and transmission systems,'' \emph{IEEE Transactions on
  Sustainable Energy}, vol.~9, no.~4, pp. 1692--1701, 2018.

\bibitem{christakou2013efficient}
K.~Christakou, J.-Y. LeBoudec, M.~Paolone, and D.-C. Tomozei, ``Efficient
  computation of sensitivity coefficients of node voltages and line currents in
  unbalanced radial electrical distribution networks,'' \emph{IEEE Trans. Smart
  Grid}, vol.~4, no.~2, pp. 741--750, 2013.

\bibitem{paolone2015optimal}
M.~Paolone, J.-Y. Le~Boudec, K.~Christakou, and D.-C. Tomozei, ``Optimal
  voltage control processes in active distribution networks,'' The Institution
  of Engineering and Technology-IET, Tech. Rep., 2015.

\bibitem{guptal2019performance}
R.~Guptal, F.~Sossan, and M.~Paolone, ``Performance assessment of linearized
  opf-based distributed real-time predictive control,'' in \emph{2019 IEEE
  Milan PowerTech}.\hskip 1em plus 0.5em minus 0.4em\relax IEEE, 2019, pp.
  1--6.

\bibitem{CEI016}
``{CEI 0-16: Reference technical rules for the connection of active and passive
  consumers to the HV and MV electrical networks of distribution Company},''
  {Italian Electrotechnical Committee}, Tech. Rep., Apr 2014.

\bibitem{pignati2015real}
M.~Pignati, M.~Popovic, S.~Barreto, R.~Cherkaoui, G.~D. Flores, J.-Y.
  Le~Boudec, M.~Mohiuddin, M.~Paolone, P.~Romano, S.~Sarri \emph{et~al.},
  ``Real-time state estimation of the epfl-campus medium-voltage grid by using
  pmus,'' in \emph{2015 IEEE Power \& Energy Society Innovative Smart Grid
  Technologies Conference (ISGT)}.\hskip 1em plus 0.5em minus 0.4em\relax IEEE,
  2015, pp. 1--5.

\bibitem{sossan2019solar}
F.~Sossan, E.~Scolari, R.~Gupta, and M.~Paolone, ``Solar irradiance estimations
  for modeling the variability of photovoltaic generation and assessing
  violations of grid constraints: A comparison between satellite and
  pyranometers measurements with load flow simulations,'' \emph{Journal of
  Renewable and Sustainable Energy}, vol.~11, no.~5, p. 056103, 2019.

\bibitem{stai_2017}
E.~Stai, L.~Reyes-Chamorro, F.~Sossan, J.~Y.~L. Boudec, and M.~Paolone,
  ``Dispatching stochastic heterogeneous resources accounting for grid and
  battery losses,'' \emph{IEEE Transactions on Smart Grid}, pp. 1--1, 2017.

\bibitem{skytte1999regulating}
K.~Skytte, ``The regulating power market on the nordic power exchange nord
  pool: an econometric analysis,'' \emph{Energy Economics}, vol.~21, no.~4, pp.
  295--308, 1999.

\bibitem{campi2008exact}
M.~C. Campi and S.~Garatti, ``The exact feasibility of randomized solutions of
  uncertain convex programs,'' \emph{SIAM Journal on Optimization}, vol.~19,
  no.~3, pp. 1211--1230, 2008.

\bibitem{bang2012existing}
C.~Bang, F.~Fock, and M.~Togeby, ``The existing nordic regulating power market:
  Flexpower wp1—report 1,'' \emph{Ea Energy Anal., Copenhagen, Denmark, Tech.
  Rep. flexpower project}, 2012.

\bibitem{ESSprices}
B.~Nykvist and M.~Nilsson, ``Rapidly falling costs of battery packs for
  electric vehicles,'' \emph{Nature Climate Change}, vol.~5, pp. 329--332, 03
  2015.

\end{thebibliography}

\vfill

\end{document}